\documentclass{iopart}

\usepackage{iopams}  
\usepackage{graphicx}

\begin{document}

\vspace*{-2.cm}

\title[Theoretical developments in heavy and light flavor energy loss]
{Theoretical developments in heavy and light flavor energy loss}

\author{I Vitev$^\S$}

\address{$^\S$Los Alamos National Laboratory, 
Theoretical Divisions, Mail Stop B283 \\ Los Alamos, NM 87544, USA \\ }


\begin{abstract}
Recent developments in the many-body perturbative QCD theory of 
inelastic parton interactions in dense nuclear matter and the  
phenomenology of strongly-interacting hard probes in heavy ion 
collisions are reviewed. We highlight the progress that has been 
made toward consistent comparison between radiative and collisional 
energy loss, the exploration of novel heavy flavor suppression 
mechanisms in the quark-gluon plasma, and the determination of the 
stopping power of cold nuclear matter. Future directions and 
opportunities for jet physics in nuclear collisions, enabled by 
the unprecedentedly high center of mass energies at the LHC, are 
also discussed. We propose that the physics of jet shapes and 
a generalizations of the well-understood inclusive particle 
suppression in the QGP will provide a new differential, and 
accurate test of the underlying QCD theory and a new precision 
tool for jet  tomography at the LHC. 
\end{abstract}

\vspace*{-.7cm}

\section{Introduction}

\begin{figure}[!t]
\begin{center}
\vspace*{+.6cm}
\includegraphics[width=7.2cm]{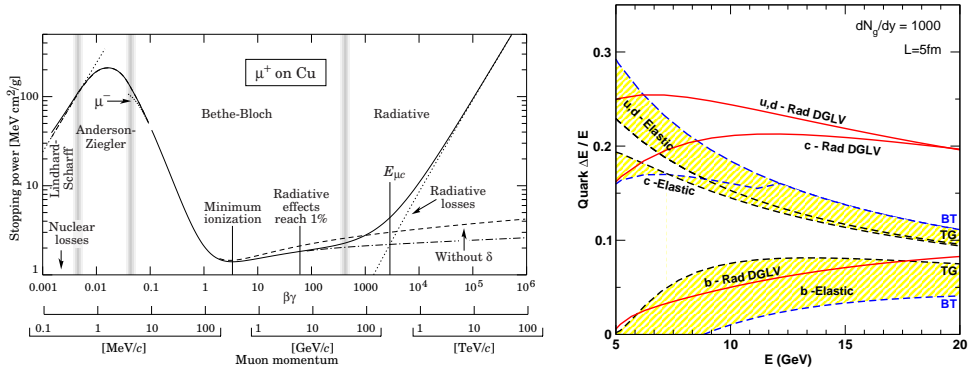}
\hspace*{.2cm} \includegraphics[width=5.2cm]{deltaEoverEnf0.eps}
\end{center}
\vspace*{-0.2cm}
\caption{ Left panel: The stopping power of Cu for $\mu^+$ versus
$ \beta \gamma $ (solid line) from Ref.~\cite{Yao:2006px}. 
Note the large-momentum  relativistic 
rise  due to radiative energy loss. Right panel: Early comparison between 
collisional and radiative energy loss in QCD from  
Ref.~\cite{Mustafa:2004dr}. }
\label{figure1}
\vspace*{-2mm}
\end{figure}

Of the interactions that charged particles undergo, as they traverse
dense matter, inelastic scattering  is undoubtedly  the most important 
and has, by far, the largest experimentally observable effect. The energy 
loss itself, $-dE/dx$, is a fundamental probe of the matter properties. 
Following the pioneering work of H. Bethe  on the stopping power of materials 
for electrons~\cite{Bethe}, precise theoretical calculations and experimental
measurements of this quantity became one of the great early successes of 
the classical and quantum theories of electromagnetic  
interactions~\cite{Yao:2006px}. 
Simplified forms of the collisional and radiative energy loss in QED 
are given by:
\begin{equation}
 \hspace*{-1.5cm} - \frac{d  E_{coll}}{dx} \approx  4 
\pi \alpha_{em}^2  z^2 Z \rho  \frac{1}{\beta^2 m}  \ln B_q \;, \; \; 
- \frac{d  E_{rad}}{dx} \approx \frac{16}{3} \alpha_{em}^3  
z^4 Z^2 \rho \frac{1}{M^2} E  \ln (\gamma \lambda) \;,  \;\; 
\label{colradQED}
\end{equation}
to illustrate their energy and path length dependence.  It is the 
consistent treatment of the fermion interactions with the medium that 
has enabled precise comparison between $ -{d E_{coll}}/{dx}$ and
$ -{d E_{rad}}/{dx}$, see the left panel in Figure~\ref{figure1}.  
It also ensures agreement between 
experimental data and the theory at the level  
of $\approx 1\%$~\cite{Yao:2006px} over six orders of magnitude in 
energy. One notes that at high energies radiative
energy loss dominates over collisional and the fermion mass dependence 
never vanishes.

In QCD, the basic results for quark and gluon stopping in matter have 
been known to exhibit the same qualitative behavior as in 
QED~\cite{Gunion:1981qs}, 
$ \Delta E_{col} \sim \alpha_s^2 L \ln E / E_c $,  $ \Delta E_{rad} 
\sim \alpha_s^3  L  E  $.  However, only after the incorporation 
of the Landau-Pomeranchuk-Migdal (LPM) type coherence effects in 
radiative energy loss calculations, for a brief overview of theoretical 
approaches see~\cite{Majumder:2007iu}, and the clear prospects of 
hard physics at RHIC and the LHC significant progress in the field of 
jet tomography has been made.

\section{Parton energy loss: heavy and light flavor suppression}

Of the experimental discoveries in nucleus-nucleus collisions 
at RHIC, the dominance of  final-state interactions 
in modifying jet and hadron production, 
when these are not close to the projectile or target fragmentation 
regions, is the most reliably 
established. It finds both theoretical support and independent 
experimental verification~\cite{Majumder:2007iu}.

\subsection{Radiative and collisional energy loss}

Jet quenching models, based on radiative energy loss, have 
been very successful in describing the observed light hadron 
attenuation in the QGP. 
In contrast, direct extrapolation of the same models to heavy $D$- 
and $B$-mesons cannot account for the measured large suppression, 
$ R^{e^\pm}_{AA} (p_T) \approx 0.25 $, of non-photonic electrons at RHIC.  
The failure of the naive extensions to explain open heavy flavor
dynamics in the QGP has lead to renewed interest in collisional 
energy loss. Indeed, in the limit of large LPM cancellation, 
for static plasmas we have~\cite{Gunion:1981qs,Majumder:2007iu}:   
\begin{equation} 
 \hspace*{-1.5cm} - \frac{d  E_{coll}}{dL} \approx   
\frac{ 2 \alpha_s }{3} \frac {\mu^2}{2} 
\log \left( \kappa \frac{T E}{ \mu^2} \right) \;, 
\;\; 
 - \frac{ d E_{rad}}{dL} \approx   
\frac{ 2 \alpha_s }{3} \frac {\mu^2 L}{\lambda_g} 
\log \left( \frac{2 E}{ \mu^2 L} \right)  \;. 
\label{colradQCD}
\end{equation}
Based on the the very similar functional dependence on the
parton $E$, early comparisons between collisional and radiative 
energy loss pointed that these can be comparable at high $p_T$ 
and in the intermediate-$p_T$ region $ -{d E_{coll}}/{dL}  
> -{d  E_{rad}}/{dL} $~\cite{Mustafa:2004dr},  see e.g. the 
right panel of Figure~\ref{figure1}.  With the results in 
Eq.~(\ref{colradQCD}) derived under different approximations 
for the interaction of the parton with the  medium, however, 
these conclusions had to be  revised. It was first shown in 
a calculation  of drag coefficients in the  QGP~\cite{Vitev:2007jj} 
that for $\gamma = E/m > {\rm few}$ radiative  energy loss 
dominates. Subsequent developments  in most theoretical
approaches to parton energy loss now appear to find similar 
results~\cite{Wang:2006qr}. Collisional energy loss should be 
included in a full description of parton propagation in matter, though
at intermediate, except for $b$ quarks, and  high $p_T$ (or $E_T$) its 
effects are small.

\subsection{Heavy flavor suppression mechanisms: from the
perturbative to the non-perturbative and back}

\begin{figure}[!t]
\vspace*{+1.cm}
\begin{center}
\includegraphics[width=5.5cm]{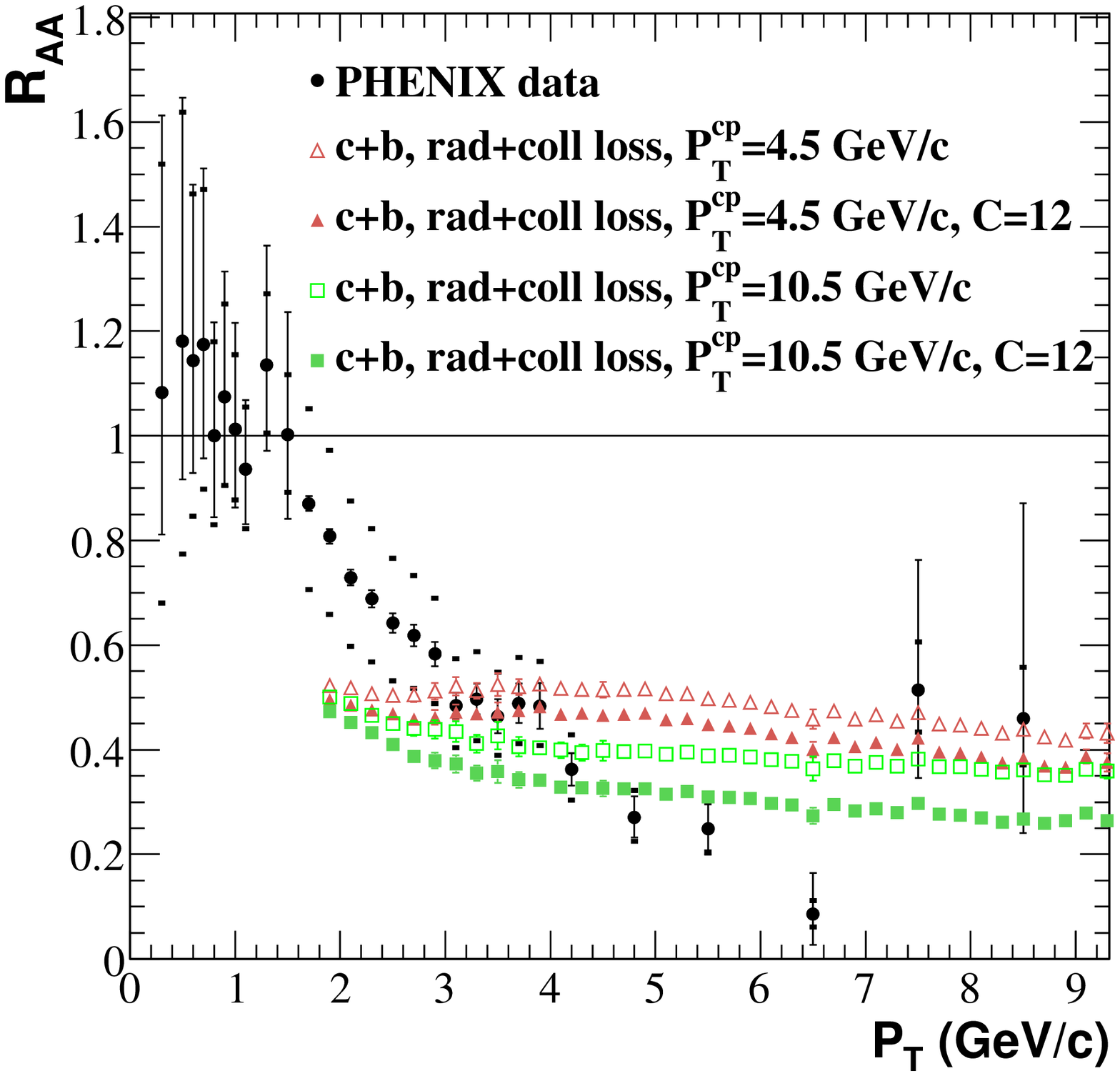}
\hspace*{.2cm}\includegraphics[width=6.9cm,height=6.5cm]{AuCuLHC.eps}
\end{center}
\vspace*{-.2cm}
\caption{ Left panel: Effects of charm baryon enhancement on the
non-photonic $e^\pm$ suppression from Ref.~\cite{MartinezGarcia:2007hf}. 
Comparison to RHIC data is shown. 
Right panel: Suppressions of $D$- and $B$-meson production 
via collisional dissociation in the QGP in central Au+Au and 
Cu+Cu reactions  at RHIC and central Pb+Pb reactions at the LHC
from Ref.~\cite{Adil:2006ra}. Note  
$R_{AA}^B \approx R_{AA}^D $ at $p_T = 10$~GeV. }
\label{figure2}
\vspace{-2mm}
\end{figure}

Compelling alternatives to partonic energy loss have been proposed
to explain the large suppression of non-photonic 
electrons at RHIC~\cite{MartinezGarcia:2007hf,Adil:2006ra}. Some of these
models rely on non-perturbative mechanisms, such as quark coalescence.  
If charm baryon production, e.g. $\Lambda_c$, is significantly 
enhanced ($C \sim 12$), the small semi-leptonic decay rate of 
$\Lambda_c$ will result in fewer electrons. The left panel of 
Figure~\ref{figure2} shows that such mechanism can largely account 
for $R^{e^\pm}_{AA}(p_T)$.

The heavy flavor ``puzzle'' has also rekindled interest in the space-time 
picture of hadronization. The short formation time of $D$- 
and $B$-mesons: 
\begin{equation} 
\hspace*{-1.5cm}\Delta y^+ \simeq \frac{1}{\Delta p^-} \; = \; 
\frac{ 2z(1-z)p^+}{  {\bf k}^2 + (1-z)m_h^2 - z(1-z)m_Q^2   } \;,  \; 
\tau_{\rm form} = \frac{ \Delta y^+}{ 1+\beta_Q} \;,
\label{tfrag}
\end{equation} 
where $\beta_Q = {p_Q}/{E_Q}$, strongly suggest that
the competing mechanisms of fragmentation and dissociation in 
the medium (inelastic processes) can emulate energy 
loss~\cite{Vitev:2007jj,Adil:2006ra}. Theoretical advances in 
relating the dissociation rate heavy meson to the QGP properties
have allowed for detailed predictions of the attenuated heavy 
favor cross sections, see the right panel of Figure~\ref{figure2}.   
A unique feature of this model is that  the large beauty meson 
mass facilitates suppression similar to that of charm mesons at
intermediate $p_T$. 
The formation time approach, formulated in ~\cite{Adil:2006ra}, 
can also be applied to evaluate the
potential of strange-quark hadronic resonances, $K^*, \phi, \Lambda^*$, to
carry information for the existence of  chiral symmetry 
restoration in the QGP state~\cite{Markert:2007gy}.

Finally, it is  possible that the suppression of heavy 
flavor is smaller than the current non-photonic electron measurements 
suggest. Experimental upgrades at RHIC, aimed at direct and separate
measurement of $D$- and $B$-meson $R_{AA}(p_T)$, and future 
measurements at the LHC will help clarify the mechanism of heavy
flavor suppression in the QGP, see e.g Table~\ref{table1}.
\begin{table}[!t]
\caption{ \label{table-heavy} Differences between models of heavy 
flavor suppression,  accessible via separate charm and beauty 
hadron measurements at $p_T \sim10$~GeV. }
\begin{center}
\begin{tabular}{@{}llll}
\br
  Model   &  \  Partonic energy 
&   \   Heavy baryon 
&  \  Collisional   \\
   &  \   loss  
&   \   enhancement 
&  \  dissociation  \\
\mr
Characteristic  & \  $R^B_{AA} \gg  R^D_{AA}$   
& \  $R^B_{AA} \gg  R^D_{AA}$ 
& \  $R^B_{AA} \approx  R^D_{AA}$   \\
features  & \  $R^{e^\pm}_{AA} >  R^{\pi,h}_{AA}$ 
& \  $R^{e^\pm}_{AA} \approx  R^{\pi,h}_{AA}$   &  
\  $R^{e^\pm}_{AA} \approx  R^{\pi,h}_{AA}$  \\
\br
\end{tabular}
\end{center}
\vspace*{-.5cm}
\label{table1}
\end{table}

\subsection{The stopping power of cold nuclear matter}

\begin{figure}[!b]
\vspace*{.cm}
\begin{center}
\includegraphics[width=6.cm,height=5.5cm]{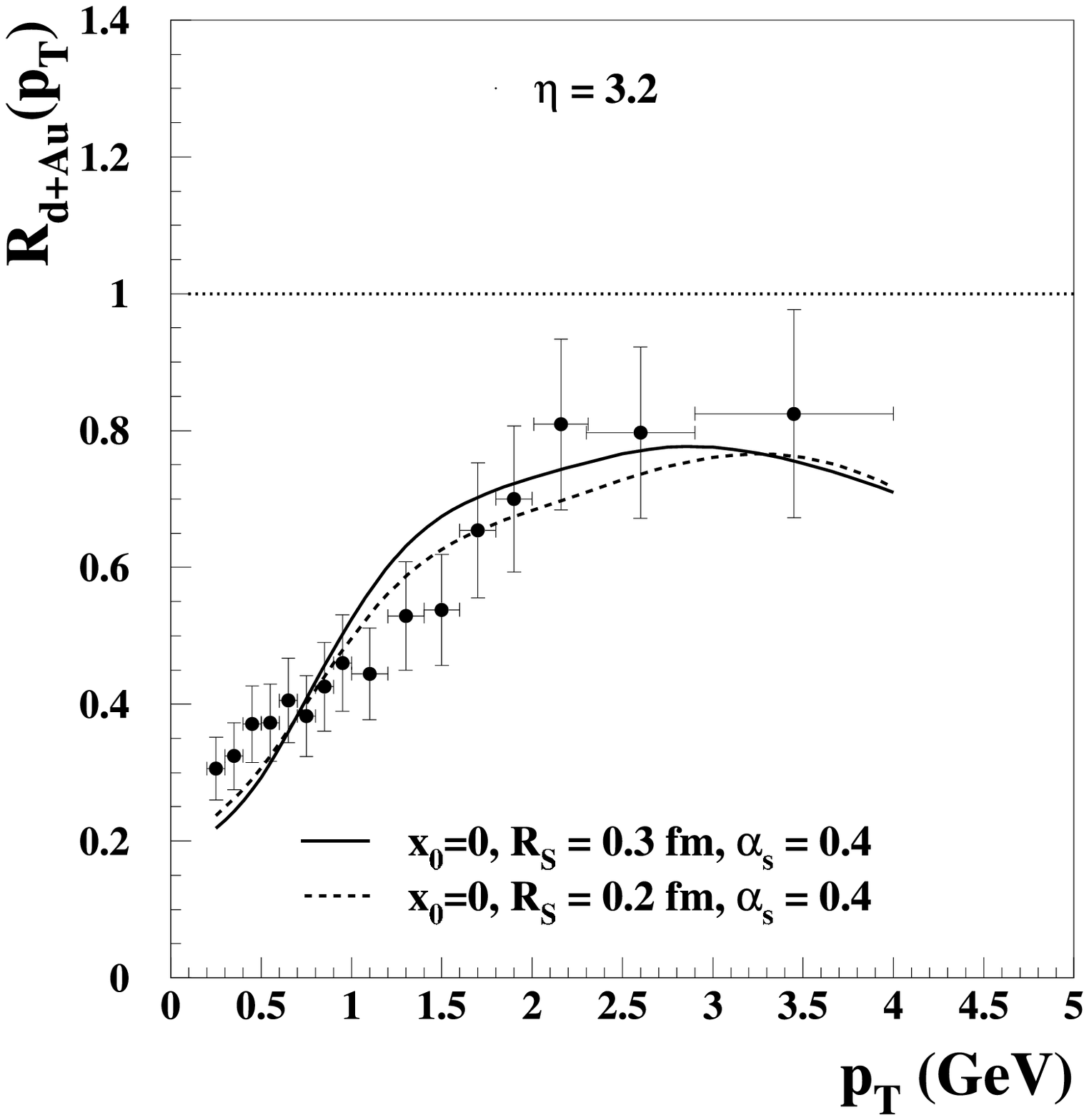}
\hspace*{.1cm}
\includegraphics[width=6.3cm]{QGloss.eps}
\end{center}
\vspace*{-0.cm}
\caption{Left panel:  Comparison of a theoretical model~\cite{Kopeliovich:2005ym} 
that includes shadowing and parton energy loss to $\eta =3.2$ $R_{dAu}(p_T)$ 
data at RHIC. Right panel: Comparison between $\Delta E_g$ and $\Delta E_q$ in 
central  collisions of large nuclei at RHIC and the LHC~\cite{Vitev:2005he} shows
large deviations from $\Delta E_g = 2.25 \Delta E_g$ for finite parton energies. }
\label{figure3}
\vspace{-2mm}
\end{figure}

In comparison to the extensive studies of
final-state parton interactions in the QGP, the 
theory of cold nuclear matter energy loss is not as well developed. 
It was realized only recently that the stopping power of large nuclei
has a sizable, possibly dominant, contribution to the suppression of 
particle production in p+A and e+A reactions and its effects are always
present in A+A reactions~\cite{Kopeliovich:2005ym,Vitev:2008vk}.

In proton-nucleus collisions, parametrizations or dynamical calculations
of nuclear shadowing cannot explain the large attenuation of hadron 
cross sections at forward rapidity~\cite{Kopeliovich:2005ym}. Combined with 
the fact that similarly large suppression is observed as a function of
Feynman $x_F = 2 p_L/ \sqrt{s}$ even at center of mass energies as low 
as $\sqrt{s} = 5$~GeV, this is a strong indication that inelastic 
parton interactions play a dominant role in altering particle production 
in p+A reactions. In this context, the significance of $\Delta E_{rad}$ 
was first emphasized in Ref.~\cite{Gavin:1991qk}  
on the example of the observed $J/\psi$ suppression in fixed target 
experiments. The left panel of Figure~\ref{figure3} shows a comparison 
of a calculation that includes nuclear shadowing and parton 
energy  loss to $R_{dAu}(p_T)$ in the forward direction at 
RHIC. Such models find  
theoretical support in a recent derivation of the stopping power of
large nuclei~\cite{Vitev:2007ve}. The LPM effect for 
initial-state radiative energy only reduces its magnitude, 
$\Delta E_{rad} \sim (\kappa_{LPM} \sim 1/6)E L$, allowing for
cold nuclear matter quenching for partons of very large 
energy $E = m_T \cosh(y-y_{\rm target})$.

    Concurrent development of ideas related to heavy ion 
phenomenology in cold and hot nuclear matter is also 
illustrated by calculations of jet conversion~\cite{Schafer:2007xh}. 
In semi-inclusive DIS this mechanism accounts qualitatively for 
the flavor dependence of the hadron suppression measured by the 
HERMES experiment. In the QGP jet  conversion, via quark and 
gluon mixing, reduces the  difference between $\Delta E_q$ and 
$\Delta E_g$.  Even though 
the rate was found to be small for realistic temperatures and 
densities, when
combined with $\Delta E_g < (C_A/C_F) \Delta E_q$ at finite 
energies~\cite{Vitev:2005he}, see the right panel of 
Figure~\ref{figure3}, it could account for the apparent 
absence of strong flavor dependence of the high-$p_T$ 
hadron quenching.

Details of the theoretical and phenomenological developments
that pertain to strongly-interacting hard probes in high 
energy  nuclear collisions are given elsewhere in these 
proceedings~\cite{contribs,posters}

\section{Jets in nuclear collisions}

There is general agreement on the physics that controls inclusive particle  
suppression in the QGP and the experimental methodology of 
determining $R_{AA}(p_T)$ is well established. Still, such measurements 
are not able, at present, to distinguish between competing theoretical 
models of energy loss. Experimental interest in multi-particle correlations 
has stimulated extensive phenomenological work to better constrain the 
mechanisms of jet-medium interactions~\cite{contribs,posters}. It appears  
that such modeling effort cannot be systematically improved due to the 
absence of factorization for the highly differential  observables~\cite{Collins:2007nk}. 
It is, therefore, critical to find alternatives that accurately reflect
the energy flow in strongly-interacting systems, have a more direct connection 
to the underlying QCD theory, and exhibit a larger discriminating power. 
We  propose~\cite{VW} that jet shapes in nuclear 
collisions and a natural generalization of leading hadron 
quenching to jets, $R_{AA}(E_T; R,p_{T\; \min} )$,  
are precisely the tools needed to leverage the expertise acquired at RHIC.

The high rate of hard probes at the LHC and the large-acceptance calorimetry, 
see e.g.~\cite{D'Enterria:2007xr}, will enable precise jet measurements. 
Discussion of the merits of and recent 
improvements in jet finding algorithms goes beyond the scope of this 
overview~\cite{Ellis:1993tq}. Understanding the QGP-induced modification 
of jet shapes is most intuitive for the cone variety, where 
$ R = \sqrt{(\eta-\eta_{\rm jet})^2 + ( \phi-\phi_{\rm jet })^2 } $ 
is the Lorentz-invariant opening angle. The differential energy 
distribution is the central quantity of interest. For 
$0\leq r \leq R$ we have: 
\begin{equation}
 \Psi(r;R) = \frac{\sum_i (E_T)_i \Theta (r-(R_{\rm jet})_i)}
{\sum_i (E_T)_i \Theta(R-(R_{\rm jet})_i)}\;, \;  
 \psi(r;R) = \frac{d\Psi(r;R)}{dr}\; .
\end{equation}

\subsection{Jet shapes in elementary p+p collisions}

\begin{figure}[!t]
\vspace*{+1.cm}
\begin{center}
\includegraphics[width=6.3cm,height=6cm]{stack-LOPCRS-rho2.eps}
\hspace*{.2cm}\includegraphics[width=6.3cm,height=6cm]{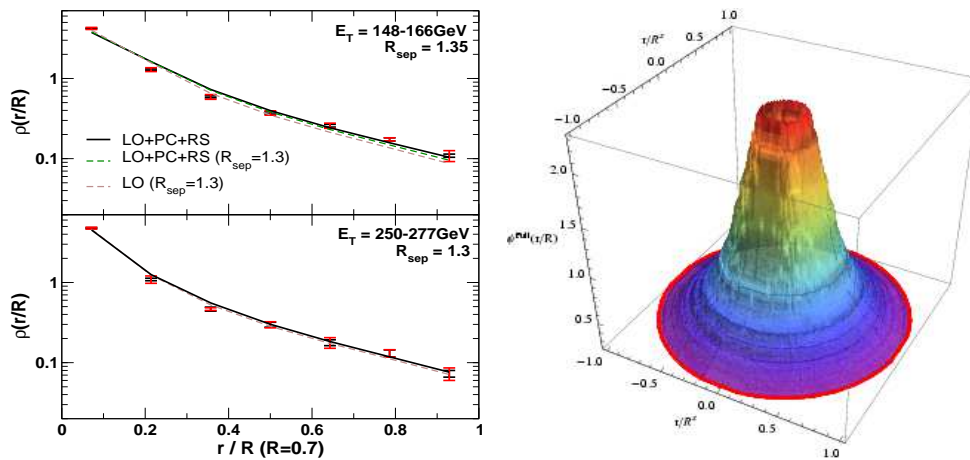}
\end{center}
\vspace*{-0.1cm}
\caption{Left panel: Jet shape calculations (R=0.7) in p+$\bar{\rm p}$ 
collisions at $\sqrt{s}=1.96$~TeV compared to the Tevatron 
data~\cite{VW}. Right panel: Preliminary results for the shape 
of a $E_T=$~50 GeV gluon  jet in $\sqrt{s}=5.5$~TeV central Pb+Pb 
collisions at the LHC~\cite{VW}. }
\label{figure4}
\vspace{-2mm}
\end{figure}

The essential features of a jets shape can be understood 
analytically~\cite{Seymour:1997kj} and arise form the 
infrared-safe QCD splitting kernel, Sudakov resummation of large 
 logarithms $ \sim \ln (r/R) $  that regulates the collinear 
divergences, initial-state radiation  present in hadronic reactions, 
power corrections that reflect the non-perturbative effects of 
hadronization, and the specifics of a jet finding algorithm:  
\begin{equation}
 \hspace*{-1.5cm} \psi(r/R) = \psi_{\rm soft}(r/R)P_{\rm Sudakov}(r/R) +
( \psi_{\rm LO }(r/R) - \psi_{\rm soft}(r/R) ) + 
 \psi_{\rm power }(r/R) \;.   \;\;
\label{psitot}
\end{equation}
Results at MLLA~\cite{VW,Seymour:1997kj} and 
Next-to-MLLA~\cite{PerezRamos:2007cr} have been compared to the Tevatron 
data on $\psi(r/R)$, e.g. see left panel of Figure~\ref{figure4}. 
Note that vacuum jets are very strongly peaked at $r \ll R$.
In elementary p+p collisions jet shapes at the LHC are very similar 
to the ones at the Tevatron, up to a different contribution of gluon 
jets, where $\langle r/R \rangle_g > \langle r/R \rangle_q $.

\subsection{Medium-induced modification of jet shapes}

Detailed derivation of the coherent inelastic parton scattering regimes
in QCD was given in~\cite{Vitev:2007ve}. In all cases, the origin of the 
LPM suppression can be tracked to the suppression or full cancellation 
of collinear,  $k_T \ll \omega$,  gluon bremsstrahlung. Heuristically, 
this can be understood as follows:
\begin{equation}
 \hspace*{-2.5cm} 
\Delta E^{rad}_{\rm LPM \; suppressed} \Rightarrow  
\frac{dI^g}{d\omega}(\omega \sim E)_{\rm LPM \; suppressed} 
 \Rightarrow  \frac{dI^g}{d\omega d^2 k_T}
(k_T \ll \omega )_{\rm LPM \; suppressed} \;, \;
\label{largeang}
\end{equation}  
and we indicate the parts of phase space where the attenuation is effective. 
The destructive quantum interference is most prominent for final-state 
radiation, where the large-angle gluon bremsstrahlung was originally  
discussed~\cite{Vitev:2005yg}. The  $k_T \rightarrow 0$  cancellation
persists to all orders in opacity~\cite{VW} and was verified numerically 
in Monte-Carlo simulations of 
${dI^g}/{d\omega d^2 k_T}$~\cite{contribs,Wicks:2008ta}.
Similar results for QGP-induced large-angle parton splitting are found  
in other approaches~\cite{Koch:2005sx}, albeit with somewhat different 
theoretical justification.

\subsection{Tomography of jets and experimental observables}

The ability to select the cone radius $R$ and the minimum 
particle energy $p_{T \; \min}$ will allow for the first time 
for a full 2D reconstruction of QGP-induced bremsstrahlung spectrum 
both in angle $r \approx k_T/\omega$ and energy $\omega$. This 
approach relies on the 
fact that the shape functions   $\psi_{\rm vac.}(r/R)$ and 
$\psi_{\rm med.}(r/R) = (1/\Delta E_{rad}) dI^g/dr$ are substantially 
different. The full medium-modified jet shape is given by:  
\begin{eqnarray}
\hspace*{-2.cm} && \hspace*{-2.cm} \psi_{\rm tot.}(r/R) = 
\frac{1}{\rm Norm}
\int_0^1 d\epsilon \,  P(\epsilon) 
\left(  \frac{1}{(1-\epsilon)^2} 
\frac{ d\sigma^{\rm pp} (R,p_{T\; \min} ) }{d^2 E_T^\prime dy} 
 \psi_{\rm vac.}(r/R)  \right.  \nonumber \\
&& \hspace*{1.3cm}
 \left. +  \; \frac{1}{ ( f(R/\infty,p_{T\; \min}/\infty) \epsilon )^2 }
  \frac{d\sigma^{\rm pp}(R,p_{T\; \min})}{d^2 E_T^{\prime \prime} dy} 
 \psi_{\rm med.}(r/R)  \right) \; . \; \quad
\label{psitotmed}
\end{eqnarray}  
In Eq.~(\ref{psitot}),  $\int_0^1  \psi_{\rm vac.,\; med.}(r/R) dr = 1$ 
and the normalization is the quenched jet cross section. 
Here $P(\epsilon)$ is the probability to lose energy due to 
multiple gluon emission and   
$f(R/\infty,p_{T\; \min}/\infty)$ is the fraction of the lost 
energy that falls within the jet cone $R$ and is carried by  
gluons of $\omega > p_{T\; \min}$.

\begin{figure}[!t]
\vspace*{+.2cm}
\begin{center}
\includegraphics[width=6.3cm]{psiVacMed-2.eps}
\hspace*{.2cm}\includegraphics[width=6.3cm,height=5.5cm]{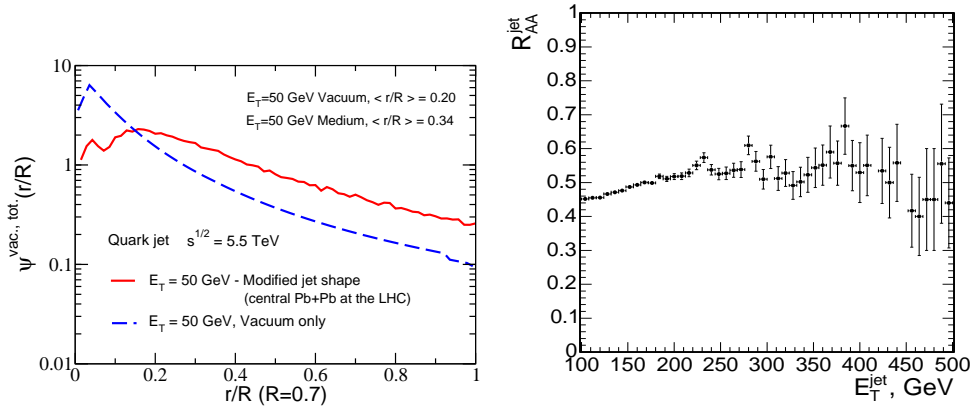}
\end{center}
\vspace*{-0.3cm}
\caption{Left panel: Preliminary comparison of the vacuum and the full 
in-medium jet shapes 
for $E_T = 50$~GeV gluon jets in central Pb+Pb collisions at the LHC 
illustrates large observable QGP effects~\cite{VW}.  
Right panel: Monte-Carlo simulation of  $R^{jet}_{AA}(E_T)$ at the LHC
for $R=0.5$ from Ref.~\cite{Lokhtin:2007ga}.   }
\label{figure5}
\vspace{-2mm}
\end{figure}

The right panel of Figure~\ref{figure4} shows preliminary results 
on the energy flow in  inclusive $E_T =50$~GeV gluon jets ($R=0.7$)  
in central Pb+Pb collisions at the LHC. Vacuum and in-medium jet
shapes differ significantly, as shown in the left panel of 
Figure~\ref{figure5}, and the mean jet opening angle 
$ \langle  r/R \rangle $ may increase by as much as 75\%.
Finally, the quenching of the jet cross sections 
$R_{AA}(E_T; R,p_{T\; \min} )$ depends critically on the  choice
or $R$  and $p_{T\; \min}$. Thus, for the same centrality, 
$E_T$  and $\sqrt{s}$  the continuum of quenching values is 
expected to help differentiate between competing models of parton 
energy loss~\cite{Majumder:2007iu}, thereby eliminating the 
order of magnitude uncertainty in the extraction of the
QGP density.  
One example of jet suppression at the LHC with a Monte-Carlo 
implementation of radiative and collisional energy 
loss~\cite{Lokhtin:2007ga} is given in the  right panel of 
Figure~\ref{figure5}. Such observables can also be easily 
generalized to photon- or $Z^0$-tagged 
jets~\cite{D'Enterria:2007xr,Mironov:2007bu}.

\section{Summary}

Recent developments in many-body perturbative 
QCD at high energies have been aimed at creating a more consistent
and detailed theory of parton and particle propagation in dense 
matter. Progress has been made in assessing the relative importance
of radiative and collisional energy loss for hard probes physics 
and first steps have been taken in studying the space-time picture
of hadronization in the medium on the example of massive quarks.   
Competing compelling models of perturbative and non-perturbative 
heavy flavor dynamics in the QGP have been proposed and will soon 
be confronted by experimental data. Parton propagation in
large nuclei and the stopping power of cold dense matter have  
become an equal and integral part of the ongoing theoretical 
developments in light of their important implications for heavy 
ion phenomenology.    

To date, uncertainties remain in the treatment of jet-medium 
interactions that cannot be resolved by the currently available 
measurements of leading particles and particle correlations 
at RHIC. We expect that the growing effort to understand 
the physics of jet shapes, jet topologies, and jet cross sections 
at the LHC will provide the possibility for qualitatively better
supported (by fundamental many-body QCD theory and numerical 
simulations) and quantitatively more precise tomography of 
nuclear matter in extremes.

\vspace*{0.2cm}\noindent\textbf{Acknowledgment:} This work is
supported by the U.S. Department of Energy  Office of Science 
under contract No.  DE-AC52-06NA25396. Helpful discussions with 
S. Wicks and B.~W.~Zhang is gratefully acknowledged.

\end{document}